\documentclass[prl,reprint,showpacs,superscriptaddress]{revtex4-2}

\usepackage{graphicx}% Include figure files
\usepackage{dcolumn}% Align table columns on decimal point
\usepackage{bm}
\usepackage{graphicx}
\usepackage{epsfig}
\usepackage{epsf}
\usepackage{amssymb}
\usepackage{amsmath,empheq,cases,dsfont}
\usepackage{amsthm}
\usepackage{mathtools}
\usepackage{enumitem}
\usepackage{multirow}
\usepackage[colorlinks=true,linkcolor=blue,citecolor=red,pdfauthor={ },pdftitle={ },pdfsubject={ },pdfkeywords={ }]{hyperref}
\usepackage{pgfplots}
\usepackage{lipsum}

\usepackage{times}

\theoremstyle{definition}

\usepackage{multirow}
\usepackage{dcolumn}
\newcolumntype{d}[1]{D{.}{.}{#1}}

\begin{document}

\title{Robust Dynamical Decoupling for the Manipulation of a Spin Network via a Single Spin}

\author{Xiaodong Yang}
\affiliation{Institute for Quantum Science and Engineering and Department of Physics, Southern University of Science and Technology, Shenzhen 518055, China}
\affiliation{Guangdong Provincial Key Laboratory of Quantum Science and Engineering,
Southern University of Science and Technology, Shenzhen 518055, China}

\author{Yunrui Ge}
\affiliation{Institute for Quantum Science and Engineering and Department of Physics, Southern University of Science and Technology, Shenzhen 518055, China}
\affiliation{Guangdong Provincial Key Laboratory of Quantum Science and Engineering,
Southern University of Science and Technology, Shenzhen 518055, China}

\author{Bo Zhang}
\email{bozhang\_quantum@bit.edu.cn}
\affiliation{Center for Quantum Technology Research, and Key Laboratory of Advanced Optoelectronic Quantum Architecture and Measurements, School of Physics, Beijing Institute of Technology, Beijing 100081, China}

\author{Jun Li}
\email{lij3@sustech.edu.cn}
\affiliation{Institute for Quantum Science and Engineering and Department of Physics, Southern University of Science and Technology, Shenzhen 518055, China}
\affiliation{Guangdong Provincial Key Laboratory of Quantum Science and Engineering,
Southern University of Science and Technology, Shenzhen 518055, China}

\begin{abstract} 
High-fidelity control of quantum systems is   crucial   for quantum information processing, but is often limited by  perturbations from   the environment  and imperfections in the applied control fields.   Here, we investigate the  combination of dynamical decoupling (DD) and robust  optimal control (ROC) to address this problem. In this combination,   ROC is employed to find robust shaped   pulses, wherein the directional derivatives of the   controlled dynamics with respect to  control errors are reduced to a desired order. Then, we incorporate ROC pulses into DD sequences, achieving    a remarkable improvement  of robustness   against multiple   error channels. We
demonstrate this method in  the example of manipulating nuclear spin bath  via an electron spin in the NV center system. Simulation results indicate that ROC based DD sequences   outperform the state-of-the-art robust DD sequences. Our work     has implications for   robust quantum control   on near-term noisy quantum devices.
\end{abstract}

\maketitle

\emph{Introduction.--}Dynamical decoupling (DD) is a well-established open-loop quantum control technique \cite{SDA16} that has  substantial applications, such as  protecting  quantum coherence  \cite{PhysRevLett.82.2417,Du09,Bollinger09,Hanson10,Zhong15},  implementing  high-fidelity quantum gates \cite{PhysRevLett.105.230503,Liu13,PhysRevLett.110.200501,PhysRevLett.115.110502}, probing noise spectrum   \cite{Oliver11,PhysRevLett.107.230501} and quantum sensing  \cite{TTW12,PhysRevX.5.041016,AMR19}. In its basic form, a DD sequence comprises  a series of  instantaneous pulses on the  system of concern, separated by   certain interpulse delays.  However, actual   pulses must have bounded strengths and hence can not be really instantaneous. Besides, there are inevitable pulse control  imperfections, including off-resonant errors and control  field fluctuations. These non-ideal  factors can cause   a DD  scheme   fail to achieve the expected performance \cite{SAS12}. To overcome this problem, a  commonly used approach is to   apply   pulses along different spatial directions to let the error of one pulse   be compensated by   other pulses \cite{AAM86,VLK03,KKV09}. Also,  one can combine DD    with  composite pulses (CP) which are robust against operational errors \cite{LMH86}. These  efforts    have yielded in the past decades a significant number of robust  DD sequences, such as  the XY8 \cite{AAM86}, CDD \cite{KKL05}, KDD  \cite{SAM11},  UR \cite{GGT17}, and random-phase \cite{WZL19} sequences. However, since  fighting against   noise effects on   current noisy intermediate-scale quantum \cite{Preskill2018quantumcomputingin} devices  remains a compelling  challenge, robust DD  continues to be a significant research problem of strong interest.

\begin{figure}[b]
\centering
\includegraphics[width=\linewidth]{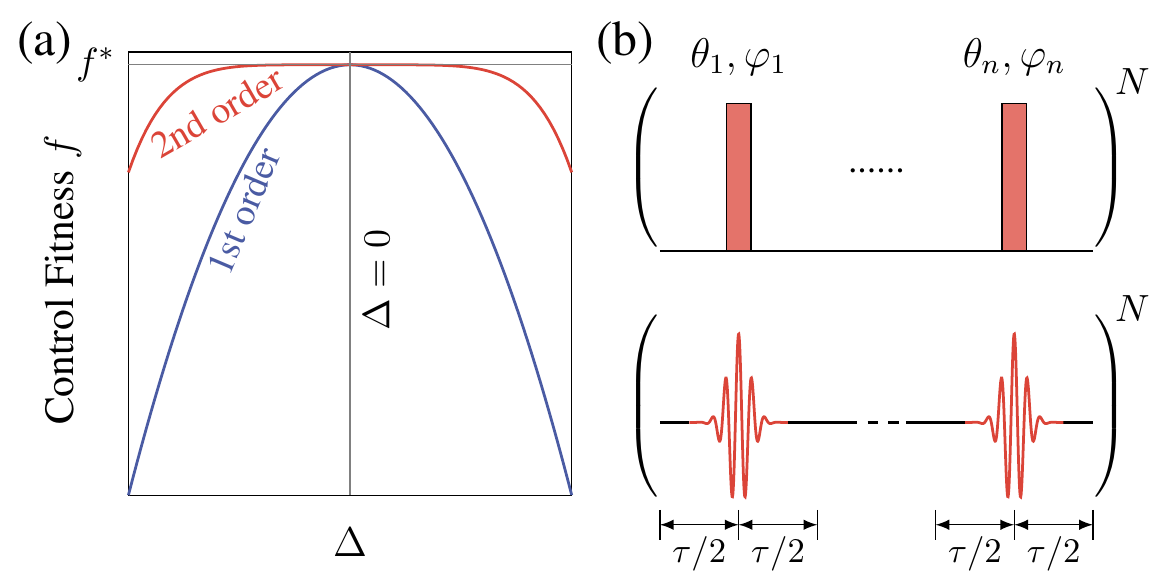}	
\caption{(a) Robustness against  pulse imperfection. The higher level to which the  derivatives of the  fitness function with respect to   the imperfections are reduced, the better. (b) Typical  DD sequence structure   combining a series of basic pulses with arbitrary shapes.}
\label{robust}
\end{figure}

Generally, for establishing a well-behaved robust DD sequence, several issues are  particularly worth considering \cite{SDA16,SAS12}. First, a satisfactory control fitness profile against   imperfections resembles that shown in Fig. \ref{robust}(a), i.e., the central robust region should be as broad and  flat as possible.  Second, it is desirable that the   DD sequence can resist several types of imperfections simultaneously. This is a natural requirement  because  in realistic situations    various perturbations  exist concurrently. Third, the   widths of the basic pulses need to be much shorter than the interpulse delays. However,     present DD sequences usually do not   meet all of these requirements. Take  CP based DD as an example,  representative CPs  like CORPSE  \cite{PhysRevA.67.042308} or BB1 \cite{BB194} correct only a single error type, and   concatenated CPs can correct simultaneously existing   errors  but only  at the price of   larger pulse-widths \cite{CCP13}, hence the    CP  technique does not easily lend itself to incorporating the  desired robustness features.

In this work, we  show that  combining  DD with robust optimal control   (ROC)   provides a    general and effective way to   enhance       robustness for   multiple pulse errors. Here, ROC is in essence quantum optimal control \cite{WG07} with  the additional robustness constraint     requiring that the leading order effects of the  imperfections are minimized. Our   basic  idea is to  replace the hard pulses in   DD   with robust   shaped pulses that are optimized via ROC   algorithms  \cite{HPZC19}.  
  The whole DD sequence is   able to exhibit  much improved error  tolerance.  
As demonstration, we apply this framework to the problem of   manipulating a spin network via a single spin. In defect-based systems, such as nitrogen-vacancy (NV) center in diamond \cite{TTW12,KSU12,LPS13} and silicon-based quantum dots \cite{PJT12,MJJ08}, a frequently encountered control scenario is to use a single probe spin to explore properties of a    quantum spin network    in the proximity. This central spin can function as a polarizer, detector, or actuator  \cite{SLO17,AAB19}, often resorting to the DD   technique. Here, in particular, we focus on the  task of nuclear spin detection and hyperpolarization by driving   the  electron spin only. Our numerical simulations show that  ROC based DD sequences outperform the best robust DD sequences available so far.

\emph{Robust DD.--}Consider a single spin as the system $S$     coupled to a spin network as its interacting environment $E$.  In the resonantly rotating   frame of $S$, the free evolution Hamiltonian is $H =  H_{SE} + H_E$. In DD, control  is acted on $S$ alone, which is implemented by a transversely applied    time-dependent field $u(t) = (u_x(t),u_y(t))$. Let $S_\alpha = \sigma_\alpha/2$ ($\alpha=x,y,z$), $\sigma_\alpha$ being the Pauli operators, then the control Hamiltonian is $H_C(u(t)) = u_x(t)S_x + u_y(t)S_y$. We   shall assume the hierarchy in the coupling strengths, $H_C \gg H_{SE} \gg H_E$. As such, we can just omit the weak term $H_E$     as it does not  appear in the DD problem to the lowest order \cite{LB13}. The general goal of DD  is to find a   suitable control sequence    to synthesize  a target effective dynamics of interest.   
%The prototypical  example is   a DD storage protocol,  where the objective is to effectively remove   $H_{SE}$. Also, DD  can be combined with multi-qubit coherent operations. For example, when the DD modulation matches certain resonance conditions, it is possible to produce an effective   Hamiltonian that  allows selective addressing or control of an individual environment spin.
However,  in the actual implementation, DD must take into account control imperfections,  as realistic  pulse profiles inevitably incorporate  pulse-shaping errors due to   finite-power and finite-bandwidth constraints in hardware. Here, we view these imperfections as perturbations and write the  perturbed Hamiltonian as 
\begin{equation}
	H(t) = H_{SE} + H_C(u(t)) + \sum_i  \epsilon_i V_i(t),    
\end{equation} 
in which  $\left\{ V_i(t)\right\}$ are noise operators, and $\left\{ \epsilon_i\right\}$ are their corresponding weights. Specifically,  we shall consider in this work mainly off-resonant and control amplitude errors, i.e., $V_1 = \Delta_{\max}S_z$ and $V_2(t) = H_C(u(t))$, where $\Delta_{\max}$ represents the maximum detuning. 
%Such errors result in   rotations with their flip angles and    phases deviate from  the ideal values, and eventually will affect the performance of   DD.
Our goal is  to devise a robust DD protocol that realizes a given control target  with sufficiently high accuracy  for a broad range of noise strengths.

We start from a typical DD sequence of the structure   shown in Fig. \ref{robust}(b). The sequence is built based on  a set of basic pulses $   \left\{ u_{\theta,0}(t): 0\le \theta \le \pi  \right\}$, here  $u_{\theta,0}(t)$ represents a control pulse    intended to effect a   single-qubit rotation $R_0(\theta)$ with   angle $\theta$ and phase  $\phi = 0$. 
More specifically, the   sequence consists of sequentially  combining $n$  pulses   from  the basic pulse set, each with a phase shift,   into a DD block, and then repeating this block for $N$ times. The $k$th pulse in the DD block is denoted as $u_k = u_{\theta_k,\phi_k}$, obtained from $u_{\theta_k,0}$ by a phase shift    $\phi_k$.   Adjacent pulses  are time separated by $\tau$, and we shall assume that for each pulse $u_{\theta_k,\phi_k}$, its   pulse-width $\tau_{u_k}$ is small compared with $\tau$.

We want  to quantify the control robustness of the considered DD sequence.  First let us look into the errors arisen in    applying   a single pulse $u_{\theta,\phi}$. The real controlled time evolution operator in the presence of perturbations is approximately $ e^{-i H_{SE} \tau_u/2} U_{\theta,\phi}  e^{-i H_{SE} \tau_u/2}$ with $U_{\theta,\phi} = \mathcal{T} e^{  -i\int_0^{\tau_u} dt (H_C(t) + \epsilon_1 V_1 + \epsilon_2 V_2(t))}$, where we only allow for the first-order effect   of $H_{SE}$  during the short period  of pulse control \cite{SupplementalMaterial}. According to the Dyson perturbative theory \cite{DFJ49},    $U_{\theta,\phi}$   can be expanded as the series $U_{\theta,\phi} = R_{\phi}(\theta) + \sum_{i_1}\epsilon_{i_1} \mathcal{D}_{U_{\theta,\phi}}^{(1)}(V_{i_1})  + \sum_{i_1,i_2} \epsilon_{i_1} \epsilon_{i_2} \mathcal{D}_{U_{\theta,\phi}}^{(2)}(V_{i_1},V_{i_2}) + \cdots$, 
where the $m$th ($m \ge 1$) order terms $\mathcal{D}_{U_{\theta,\phi}}^{(m)}(V_{i_1}, ..., V_{i_m}) = (-i)^m R_{\phi}(\theta) \int_0^{\tau_u} dt_1  \cdots  \int_0^{t_{m-1}} dt_m  \widetilde V_{i_1}(t_1) \cdots \widetilde V_{i_m}(t_m)$ for  $i_1, ..., i_m = 1,2$  are referred to as the $m$th order directional derivatives \cite{HPZC19}, characterizing   the perturbative effects due to the noise generators $\left\{ V_i \right\}$.  
Here, $\widetilde V_i(t) = U^\dag_C(t)  V_i(t) U_C(t)$ with $U_C(t) =  \mathcal{T} e^{  -i\int_0^{t} ds  H_C(s)}$.    For any $m$ and for any time-ordered operators $V_{i_1}, ..., V_{i_m}$, we can derive that \cite{SupplementalMaterial}
\begin{equation}
	\mathcal{D}_{U_{\theta,\phi}}^{(m)} = e^{-i\phi S_z} \mathcal{D}_{U_{\theta,0}}^{(m)} e^{i\phi S_z}.
	\label{error}
\end{equation} 
%In the derivation, we have used the following facts: $e^{-i\phi S_z} S_z e^{i\phi S_z} = S_z$, and $e^{i\phi S_z} H_C(u_{\theta,\phi}(t)) e^{-i\phi S_z} = H_C(u_{\theta,0}(t))$; 
Thus there is $\left\| \mathcal{D}_{U_{\theta,\phi}}^{(m)} \right\|^2 = \left\| \mathcal{D}_{U_{\theta,0}}^{(m)} \right\|^2$,    implying that we need only to consider the robustness of the basic pulses. 

Now, for the whole DD block, the real propagator goes   
\begin{equation}
	U_\text{DD} = W_n U_{\theta_n,\phi_n} W_n    \cdots W_1 U_{\theta_1,\phi_1} W_1,
	\label{block_error} 
\end{equation} 
where $W_k = e^{-i (H_{SE} \tau/2 + \epsilon_1 V_1 (\tau - \tau_{u_k})/2)}$ is the free evolution operator which has incorporated   the $H_{SE}$ evolution  \cite{SupplementalMaterial}. 
%Now, because that (i) Eq. (\ref{block_error}) and Eq. (\ref{pulse_error}) tell us  that the   deviation of $U_\text{DD}$ from the ideal target is captured by the set of error terms $U_{\theta_k,\phi_k}^{(m)}$, and (ii) Eq. (\ref{error}) implies that $\left\| U_{\theta_k,\phi_k}^{(m)} \right\| = \left\| U_{\theta_k,0}^{(m)} \right\|$, we can say that, enhancing the  robustness  of the basic pulses $u_{\theta,0}(t)$  is  crucial   for the robustness of the DD sequence.
It is then clear that the errors in $U_{DD}$ in Eq. (\ref{block_error}) include detuning errors in   $W_k$ as well as both detuning and control amplitude errors in   $U_{\theta_k,\phi_k}$, up to first-order approximation with respect to $H_{SE}$. Hence,    there are two feasible  approaches     toward  robust DD. On the one hand, 
%However, as we have explained, the always existing pulse imperfections    in practical implementations deteriorate the    control performance.
DD seeks to use  a properly designed phase modulation strategy  so  that the errors   will cancel rather than accumulate.  Notable examples include the two-axis XY-family  \cite{AAM86}, the multi-axis UR  \cite{GGT17}, and the random-phase \cite{WZL19} or correlated-random-phase \cite{WZC20}  sequences.  On the other hand, by minimizing the magnitudes of the directional derivatives $\mathcal{D}_{U_{\theta_k,0}}^{(m)}$, the robustness of the pulses $u_{\theta_k,\phi_k}(t)$ are enhanced, and  the total error of the full sequence will be accordingly decreased. Furthermore, if both approaches are taken simultaneously, we anticipate  that the DD robustness property  will get improved to an even larger degree.

\begin{figure*}
\centering
	\includegraphics[width=0.9\textwidth]{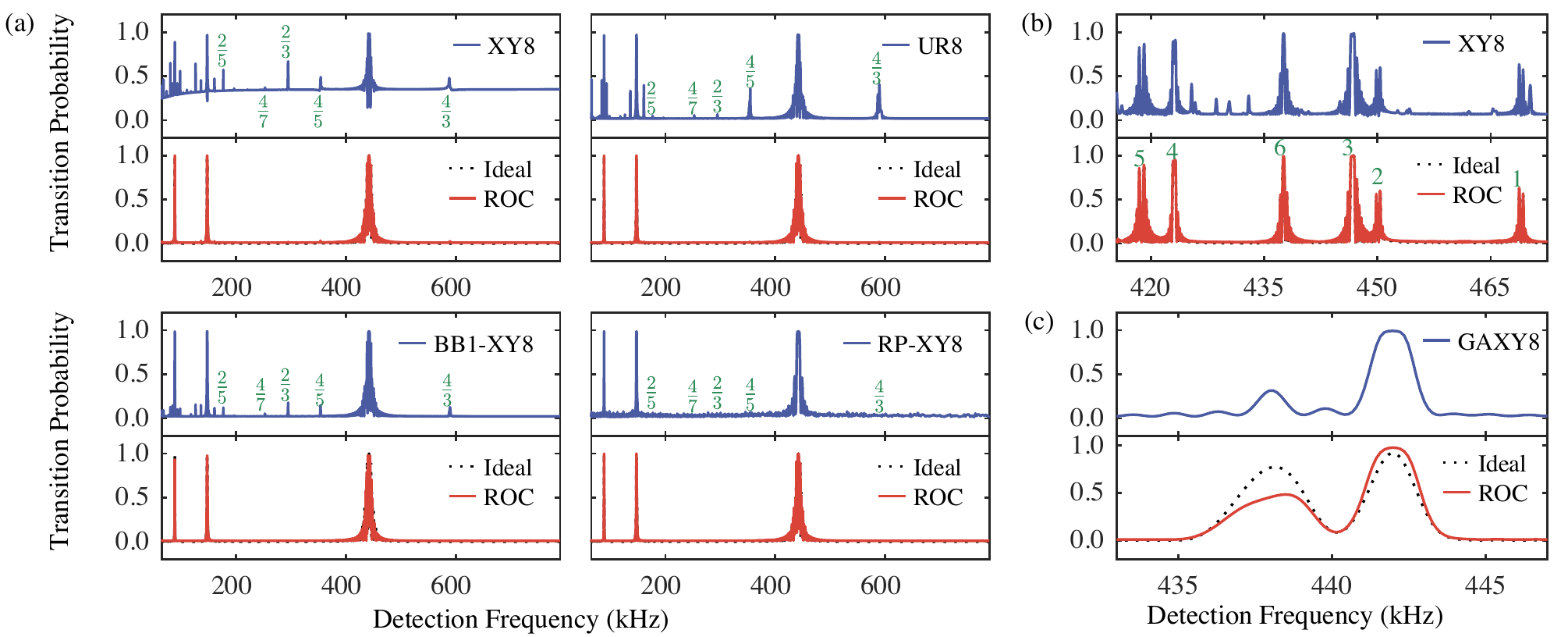}
	\caption{ Simulated spectra of detecting nuclear spins, where the   control imperfection parameters   are set as detuning $ 8 \% \times \Omega_{\max}$, amplitude error $8\% \times\Omega_{\max}$, and     $\pi$ pulse with a limit width      $1/(2\Omega_{\max}) =20~$ns. The number of cycles  for (a), (b), and (c) are 96, 32, and 64, respectively. (a) Signal from a single spin with $\omega_I=2\pi \times 441.91~$kHz detected by various DD sequences. The results for   XY8, UR8,  BB1-XY8,   RP-XY8 and their ROC counterparts are presented. Clearly,
using our ROC pulses excellently solves the   baseline distortion problem and eliminates the spurious peaks (green marker) that are located at $2\omega_I/k$ and $4\omega_I/k$ ($k$ is odd)    \cite{LMB15}.   (b) Signal from six spins (see simulation parameters in \cite{SupplementalMaterial}). The   miscellaneous peaks appeared in applying XY8 greatly hamper the identification of each spin, but can be circumvented by  ROC-XY8. (c) Signal from two spectrally close spins with $\omega_I^1=2\pi \times 441.91~$kHz and $\omega_I^2=2\pi \times 437.53~$kHz. It can be seen that, the ROC based sequence shows better robustness against imperfections in resolving the two spins.  }
		\label{spindetect}
\end{figure*}

A particularly effective approach for
devising shaped pulses that are robust to several simultaneously existing perturbations is the ROC algorithm developed in Ref. \cite{HPZC19}. In our subsequent numerical simulations, we shall generally follow this approach. Basically, we attempt to solve the following    multi-objective optimization problem: for a given target rotation angle $\theta$, to optimize the shape of the pulse $u = u_{\theta,0}(t)$   to (i) maximize the fidelity between the unperturbed propagator and the target rotation $R_0^\dag(\theta)$, and meanwhile (ii) minimize  $\left\| \mathcal{D}_{U_{\theta_k,0}}^{(m)} \right\|^2$ for $m$ up to a certain order.
One can  employ either the known gradient ascent pulse engineering algorithm \cite{NTC05} or gradient-free methods like the differential evolution \cite{PPP17,DDX20,ZEG15} algorithm  to perform the pulse search.
Note that in the entire space of admissible control functions, there may exist plenty of solutions that can accomplish the same control goal. However, as a realistic pulse generator has limited bandwidth, we shall exclude   consideration of    abruptly changing   pulse  shapes. We put the algorithm implementation details in     Supplementary Materials  \cite{SupplementalMaterial}. In the next section, we study the applications of  our optimized robust   pulses.

\emph{Applications.--}Consider the system of an electron spin $S$ ($S=1/2$) of an NV center in diamond and surrounding   nuclear spins $I$  ($I=1/2$). The system is under a static magnetic field    applied along
the NV axis which is, by convention, defined as the $\hat z$ axis.
The electron spin is subject to intermediately applied microwave control field  $u(t)$. In a rotating frame with respect to $\Omega_S S_z$,   the Hamiltonian after the secular approximation reads  
\begin{equation}
	H    =    \sum_{j}  \Omega_I I^j_z +  \sum_{j} S_z(A^j_{zz} I_z^j + A^j_{zx}   I_x^j)  + H_C(u(t)), \nonumber
\end{equation} 
in which $\Omega_S$ ($\Omega_I$) denotes the electron (nuclear) Larmor frequency,  $A_{zz}$ and $A_{zx}$ are the hyperfine coupling strengths.
In our numerical studies, we assume the existence of a minimum switching time $t_{\min} = 1$ ns for pulse modulation and   require all control amplitudes to be bounded by $\Omega_{\max} = 2\pi \times 25$ MHz. 
Three common pulse imperfections, namely detuning, control amplitude error,    and finite pulse-width, are under consideration. The details of pulse optimization  can be found in  Supplementary Materials \cite{SupplementalMaterial}.

First, we investigate the detection of nuclear spins via the central electron spin.   Nuclear spins exist naturally in abundance in   NV center systems,   providing   possible  additional   quantum resources  \cite{LTD10,PhysRevX.9.031045,AMR19}. A prerequisite to exploit this potential is to detect and characterize them, usually with the aid of the electron spin.  Ideally,  the $j$th nuclear spin with the effective Larmor frequency $\omega_I^j=\Omega_I+A_{zz}^j/2$ can induce a sharp electron signal in the observed spectrum,  by applying a multi-pulse DD sequence on the electron spin with the interpulse delay $\tau=k\pi/\omega_I^j$ ($k$ is odd) \cite{TTW12}.  Nevertheless, addressing and controlling nuclear spins is challenging as the spins are generally embedded in noisy spin bath, and the inevitable pulse imperfections can induce spurious peaks \cite{LMB15} and distortion of the spectrum \cite{CJW15}.  To avoid false identification, many robust DD sequences were proposed \cite{GGT17,WZL19,SAM11,CJW15,CJZ16}. However, when the pulse imperfections are relatively large and exist simultaneously, we find that the performance of the present robust DD sequences can be further improved by using ROC pulses.

As an initial demonstration, let us consider the situation of detecting a single    $^{13}$C nuclear spin. The simulated spectra with using four representative robust DD sequences are shown in Fig. \ref{spindetect}(a). For conventional XY8 sequence,  there exist miscellaneous errors      in the signal   including,   the tilted baseline that causes severely reduced contrast, the phase-distorted   resonant peaks, and the   spurious peaks appearing due to  finite pulse-width and detuning effects \cite{LMB15}, so   the spectrum is hard to analyze and identify.   Substituting the square-shaped $\pi$ pulses with   BB1 in XY8  (BB1-XY8)  \cite{BB194}, or using the UR8 sequence  \cite{GGT17},     improves the signal, but still does not solve the spurious peak problem. A recently proposed random-phase XY8 (RP-XY8) sequence \cite{WZL19} can largely suppress   the spurious peaks and maintain the signal contrast, but still induce dense small miscellaneous peaks near the baseline. It is found that,   for all these sequences, if implemented with  our ROC pulses ($t_{\text{ROP}}=80~$ns), then the errors almost disappear, giving a rather clean spectrum   close to the ideal one.

%Using the UR8 sequence \cite{GGT17}, the baseline shift is eliminated but the spurious peaks are even larger. The robustness of this sequence can also be improved by using our ROP; see the top right panel of Fig. \ref{spindetect}(a). By substituting the $\pi$ pulses with the well-known BB1 composite pulse ($t_{\text{BB1}}=100~$ns) \cite{LMH86}, the resultant BXY8 sequence can only slightly suppress the spurious peaks.  Replacing the basic pulses in BXY8 with ROP also enhances its robustness close to the ideal case; see the bottom left panel of Fig. \ref{spindetect}(a).  The recent proposed random-phase XY8 sequence (marked as RPXY8) \cite{WZL19} can almost suppress all the spurious peaks and maintain the contract ratio, but induce dense small miscellaneous peaks near the baseline, as shown in the bottom right panel of Fig. \ref{spindetect}(a). This can be further removed by using our ROP. See simulations details in Supplementary Materials \cite{SupplementalMaterial}.

Next, we consider detecting multiple $^{13}$C nuclear spins. The case of six nuclear spins is presented in Fig. \ref{spindetect}(b). As can be seen from the spectrum, a bunch of   spurious peaks exist, making it a difficult task to identify the spins. This gets even worse when the cycle number $N$ is large. Again, combining  XY8 with our ROC pulses    suppresses all the unwanted peaks to a satisfactory extent. As an additional demonstration example, we have   considered the problem of detecting  two spectrally close   nuclear spins. This situation  demands good selectivity of the control sequence in addressing the nuclear spins individually. Recently, Ref. \cite{HJW18} proposed an effective soft temporal modulation sequence  named Gaussian AXY8, which has better selectivity than conventional XY8. However, when taking into account of pulse imperfections, distortions occur in the transition probabilities of the spins, leading to off-resonant side peaks and hence  reducing   the signal contrast, as shown in Fig. \ref{spindetect}(c).   We thus test with our  ROC pulses, finding that the side peaks around the spin resonances are all removed, and consequently the two spins are well resolved; see   details in Supplementary Materials \cite{SupplementalMaterial}. Clearly, our approach is of great benefit when fitting dense signals for many close nuclear spins, especially   when spurious resonances exist. 

\begin{figure}[t]
\centering
	\includegraphics[width=0.95\linewidth]{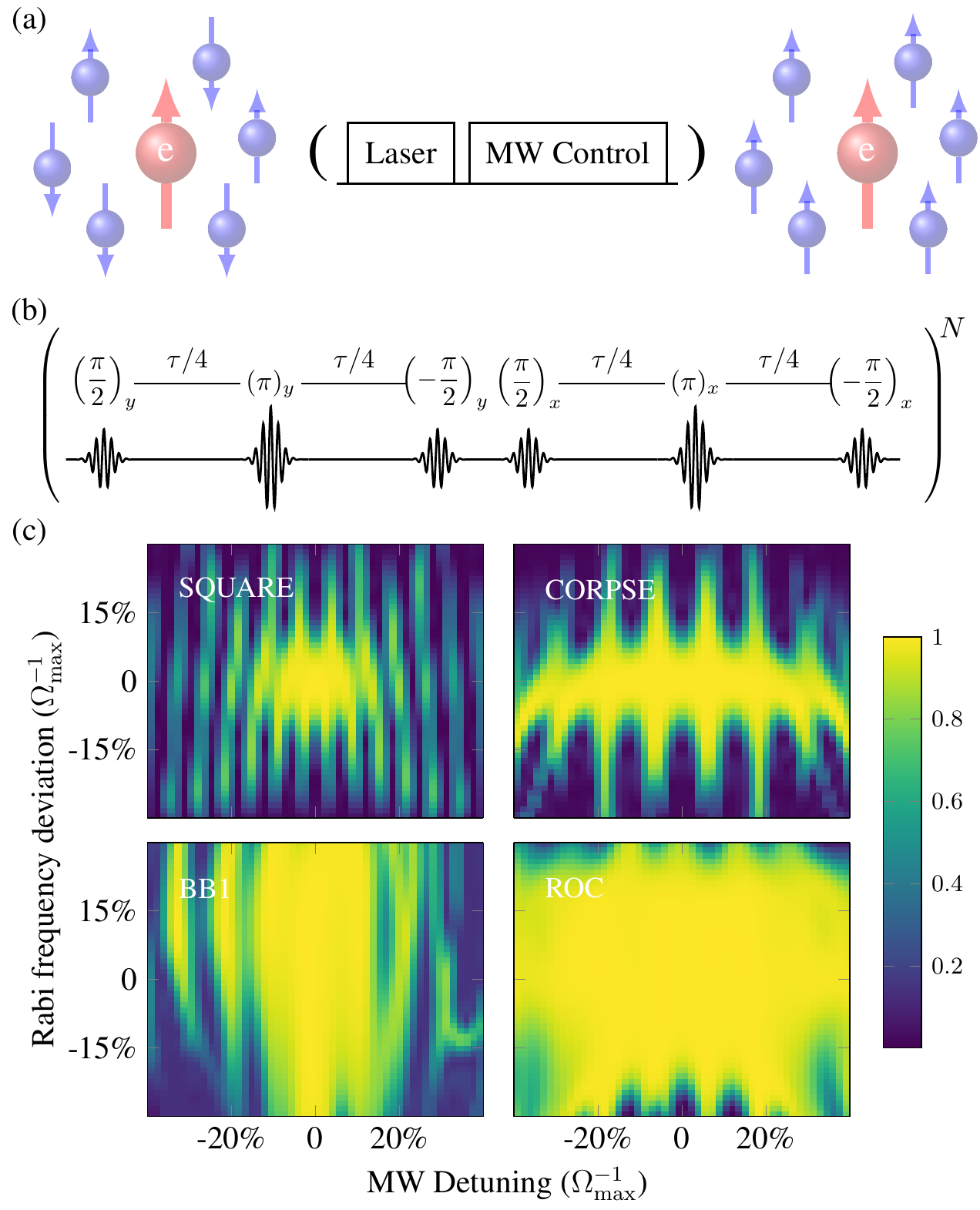}
	\caption{(a) Schematics of the  microwave (MW) pulse sequence on the electron spin for pulsed   polarization transfer to the surrounding nuclear spins.   (b) The PulsePol sequence.  As stated, the pulse duration should not take a large fraction of the free evolution time.   (c) Comparison of error-resistance of four different PulsePol sequences.}
	\label{dnp}
\end{figure}

Now, we turn to the application to  dynamical nuclear polarization (DNP) of the nuclear spins. DNP aims to transfer the almost perfect polarization of the   optically initialized    electron spin   to its surrounding nuclear spins; see Fig. \ref{dnp}(a). This could result  in an enhancement of the nuclear signal by several orders of magnitude, which is crucial for a variety of quantum technologies such as  magnetic resonance imaging \cite{ERT15}. The goal of achieving high-efficiency DNP in NV centers in diamond  has  attracted a lot of   interests in recent years but remains   technically challenging. Key factors affecting effective polarization include: the   large spectral random orientation of the NV center axes and Rabi power imperfection.  For example,   conventionally DNP   uses the Hartmann-Hahn  resonance \cite{SJS17}, the corresponding NOVEL sequence \cite{HDSW88} works     only in a narrow frequency range. Hence, there have been efforts made in devising more robust DNP protocols  \cite{CQS15,JSI16,SJS17}.  A significant advance was made by Ref. \cite{ISJ18}, which put forward a Hamiltonian engineering based approach termed PulsePol, featuring both fast polarization and excellent robustness. Yet, we find that  the results presented in Ref. \cite{ISJ18} can still be  improved, with using our ROC shaped pulses.

For simplicity, we consider a single electron coupled to a single nuclear spin. The PulsePol sequence, as shown in Fig. \ref{dnp}(b),  consists of $N$ cycles  each with four $\pi/2$-pulses and two $\pi$-pulses,  and operates for a choice of pulse spacing $\tau = l\pi/\omega_I$ for odd $l$.   Ideally, it  produces   an effective flip-flop Hamiltonian as $H_\text{eff} =   \alpha A_{zx} (S_+ I_- + S_- I_+)$, where $S_{\pm} = S_x \pm iS_y$, $I_{\pm} = I_x \pm i I_y$.   
As such, full polarization transfer   is achieved at time $2N\tau = 1/(\alpha A_{zx})$. However, because real pulses take a finite time, the free evolutions between the pulses need be adjusted according to $\tau_\text{free} = \tau - 4 t_{\pi/2} - 2 t_\pi $, where $t_{\pi/2}$ and $t_\pi$ denote the time length of the $\pi/2$ and $\pi$-pulses. As a consequence, the number of cycles $N$ required for full polarization transfer varies depending on the specific pulses used. In our simulations, we compare the robustness  performance   achieved from using four different realizations of the   rotations, namely squared, CORPSE, BB1, and our ROC pulses. The simulations are implemented with the coupling strength  $A_{zx} = 0.03$ MHz, the resonance $l=5$, and the same cycle number $N=29$. The results are shown in Fig. \ref{dnp}(c).  First, we clearly see that the robustness of PulsePol for using rectangular pulses is the worst. Then, the performance can be    improved against detuning errors for CORPSE or against  Rabi frequency errors   for BB1. Finally, best results are achieved with our ROC pulses. There is a rather large flat   central area of the robustness profile: the flat region satisfying   $f \ge 0.99$ covers the entire error range of $\pm 5$ MHz detuning   and  $\pm 10\%$ Rabi frequency deviation.

\emph{Conclusion.--}We have presented a  flexible  and effective framework for   integrating   DD with ROC. Unlike   hard pulses, shaped pulses produced by ROC with smooth constraints are   friendly to hardware implementation.  In principle,  ROC pulses can be incorporated into any DD sequence for       robustness improvement.  
This is confirmed in our test   examples of nuclear spin bath detection and hyperpolarization on NV  center systems.   Our method is ready to be exploited in many other    spin network related   problems, such as creating collective quantum memory \cite{PhysRevLett.123.140502}, probing   thermalization \cite{PDZ20},  exploring condensed matter phenomena \cite{LIS17} and studying magnetic dynamics of nanoparticles \cite{SDH15}.
ROC based DD sequences can resist    multiple  errors simultaneously. This  is especially important for near-term noisy quantum devices. For instance,   recent benchmark experiments in  superconducting qubits have identified   real coexisting error channels as well as       their respective contribution \cite{Martinis14,Reagoreaao3603}, which are to be compensated for realization of high-fidelity quantum computing. 
Therefore, a meaningful extension of the current work is to explore how to achieve best robustness given knowledge of     error channels and their   weights. 
Lastly  and rather importantly, future research could   explore novel ROC strategies that can deal with time-dependent   perturbations, which is a key ingredient to allowing robust DD to find a broader range of application scenarios     \cite{SAB14, Plenio18}.

\emph{Acknowledgments}. This work was supported by   the National Natural Science Foundation of China (Grants No. 11975117, No. 92065111 and No. 12004037),  and Guangdong Provincial Key Laboratory (Grant No. 2019B121203002).

% \bibliography{manuscript} 


\newpage
\onecolumngrid

\vspace{100pt}
\begin{center}
{\large\bfseries Supplementary Material}	
\end{center}

\section{Analysis of the Effects of Control Imperfections on DD}
In the resonantly rotating reference frame of $S$, the controlled system has the following Hamiltonian
\begin{equation}
	H(t) =   H_E + H_{SE} + H_C(u(t)) + \epsilon_1 V_1  +   \epsilon_2 V_2(t).
\end{equation}
Here, the meaning of each term is   as   described in the main text. For the sake of simplicity  $H_E$ will not be written explicitly.
The real propagator of the DD block in the presence of perturbations is written as
\begin{equation}
	U_\text{DD} = W_n U_{\theta_n,\phi_n} W_n    \cdots W_1 U_{\theta_1,\phi_1} W_1,
	\label{block_error} 
\end{equation} 
where $W_k = e^{-i (H_{SE} + \epsilon_1 V_1)(\tau - \tau_{u_k})/2}$ is the free evolution propagator, and $U_{\theta_k,\phi_k} = \mathcal{T} e^{  -i\int_0^{\tau_{u_k}} dt H(t)  }$ is the real propagator for  the $k$th pulse $u_{\theta_k,\phi_k}(t)$. Because $H_{SE}$ is   small   in terms of the short pulse-width $\tau_{u_k}$, then   to first order approximation, there is $ U_{\theta_k,\phi_k} = e^{-i H_{SE} \tau_{u_k}/2} \mathcal{T} \exp \left( -i\int_0^{\tau_{u_k}} dt \left[H_C(u(t)) +  \epsilon_1 V_1  +   \epsilon_2 V_2(t) \right] \right) e^{-i H_{SE} \tau_{u_k}/2}.$ 
The $H_{SE}$ evolutions    can be incorporated into the free evolution operators $W_k$, so they   evolve  for the right amount of time. Accordingly, we only need to analyze
\[ U_{\theta_k,\phi_k} =  \mathcal{T} \exp \left( -i\int_0^{\tau_{u_k}} dt \left[H_C(u_{\theta_k,\phi_k}(t)) +  \epsilon_1 V_1  +   \epsilon_2 V_2(t) \right] \right), \]
which we do in the following.
 
First, for pulse $u_{\theta,\phi}(t)$, in the absence of perturbations, we denote the ideal time-evolution operator by
\begin{equation}
	U^\text{ideal}_{\theta,\phi}(t) =  \mathcal{T} \exp \left\{-i \int_0^{\tau_u} dt   H_C(u_{\theta,\phi}(t))  \right\},    \nonumber
\end{equation}
Because $u_{\theta,\phi}$ is just $u_{\theta,0}$ phase shifted by $\phi$, so $H_C(u_{\theta,\phi}(t)) = e^{-i\phi S_z } H_C(u_{\theta,0}(t)) e^{i\phi S_z }$.
There are: (i) for any time $t$, $U^\text{ideal}_{\theta,\phi}(t) = e^{-i\phi S_z } U^\text{ideal}_{\theta,0}(t) e^{i\phi S_z }$ and (ii) $U^\text{ideal}_{\theta,\phi}(\tau_u) = R_{\phi}(\theta)$.

Now consider the presence of perturbations.
To separate out the deviation of the real propagator $U_{\theta,\phi}(t)$  from the ideal propagator $U^\text{ideal}_{\theta,\phi}(t)$ caused by    $V_1$ and    $V_2(t)$, it is convenient to work in the toggling frame, i.e., $U_{\theta,\phi}(t) = U_C(t) U_\text{tog}(t)$, where   $U_C(t) =  \mathcal{T} e^{  -i\int_0^{t} ds  H_C(u_{\theta,\phi}(s))}$, and the toggling frame propagator $U_\text{tog}(t) = \mathcal{T} e^{  -i\int_0^{t} ds  (\epsilon_1\widetilde V_1(s) + \epsilon_2\widetilde V_2(s))}$ with $\widetilde V_i(s) =  U^\dag_C(s)   V_i(s) U_C(s)$ can be expanded with the perturbative operators. This leads to the Dyson series
\begin{equation}
	U^\text{real}_{\theta,\phi}(\tau_u)   = R_{\phi}(\theta) + \sum_{i_1}\epsilon_{i_1} \mathcal{D}_{U_{\theta,\phi}}^{(1)}(V_{i_1})  + \sum_{i_1,i_2} \epsilon_{i_1} \epsilon_{i_2} \mathcal{D}_{U_{\theta,\phi}}^{(2)}(V_{i_1},V_{i_2}) + \cdots,
\end{equation}
where the directional derivatives have the following expression
\begin{equation}
	\mathcal{D}_{U_{\theta,\phi}}^{(m)}(V_{i_1}, ..., V_{i_m}) = (-i)^m R_{\phi}(\theta) \int_0^{\tau_u} dt_1  \cdots  \int_0^{t_{m-1}} dt_m  \widetilde V_{i_1}(t_1) \cdots \widetilde V_{i_m}(t_m).
\end{equation}
Because that
\begin{align}
	\widetilde V_1(t) & = {U^\text{ideal}_{\theta,\phi}}^\dag (t) S_z U^\text{ideal}_{\theta,\phi}(t) = e^{-i\phi S_z } {U^\text{ideal}_{\theta,0}}^\dag (t) S_z U^\text{ideal}_{\theta,0}(t) e^{i\phi S_z }, \nonumber \\
	\widetilde V_2(t) & = {U^\text{ideal}_{\theta,\phi}}^\dag (t) H_C(u_{\theta,\phi}(t)) U^\text{ideal}_{\theta,\phi}(t) = e^{-i\phi S_z } {U^\text{ideal}_{\theta,0}}^\dag (t) H_C(u_{\theta,0}(t)) U^\text{ideal}_{\theta,0}(t) e^{i\phi S_z },  \nonumber
\end{align}
this implies that
\begin{equation}
	\mathcal{D}_{U_{\theta,\phi}}^{(m)}(V_{i_1},...,V_{i_m}) = e^{-i\phi S_z} \mathcal{D}_{U_{\theta,0}}^{(m)}(V_{i_1},...,V_{i_m}) e^{i\phi S_z}. \nonumber
\end{equation}
where $\mathcal{D}_{U_{\theta,0}}^{(m)}(V_{i_1},...,V_{i_m})$ are the directional derivatives for the pulse $u_{\theta,0}(t)$.
Furthermore, we  have
$	\left\| \mathcal{D}_{U_{\theta,\phi}}^{(m)} \right\|^2 = \left\| \mathcal{D}_{U_{\theta,0}}^{(m)} \right\|^2$.
As our considered DD sequence is just   $u_{\theta,0}$ combined with a phase modulation strategy, we deduce  that  we only need to consider the robustness of the basic pulses $u_{\theta,0}$. In the next section, we describe   how to optimize these pulses, mainly focusing on $x$-rotational $\theta=\pi/2$ and $\theta=\pi$ pulses.

\section{Robust Optimal Control}

\subsection{Van Loan Block Differential Equation Formalism}
The robust optimal control method that we adopt in our work is due to Ref. \cite{HPZC19}.  

Let $V_1 =\Delta_{\max} S_z$ and $V_2 = H_C(u(t))$. We set $\Delta_{\max}=\Omega_{\max}=25~$MHz, $\epsilon_1 = 40\%$ and $\epsilon_2 = 30\%$ in all our subsequent numerical studies. The Van Loan block differential equations are constructed as follows ($i_1,i_2 = $ 1 or 2)
	\begin{equation}
	\dot V(t) = -i \left( \begin{matrix} 
  H_C(u(t)) & V_{i_1}(t) & 0 \\
 0 & H_C(u(t))   & V_{i_2}(t) \\
 0 & 0 & H_C(u(t))
\end{matrix}
\right)V(t).
\end{equation}
Its time-evolution solution is, according to Van Loan integral formula, given by
\begin{equation}
	V(\tau_u)   = \mathcal{T} \exp \left[ -i \int_0^{\tau_u} dt   \left( \begin{matrix} 
  H_C(u(t)) & V_{i_1}(t) & 0 \\
 0 & H_C(u(t))   & V_{i_2}(t) \\
 0 & 0 & H_C(u(t))
\end{matrix}
\right) \right]   
  = \left( \begin{matrix} 
 U_C(\tau_u) & \mathcal{D}^{(1)}_{U_C} (V_{i_1}) & \mathcal{D}^{(2)}_{U_C} (V_{i_1},V_{i_2})\\
 0 & U_C(\tau_u)   & \mathcal{D}^{(1)}_{U_C} (V_{i_2}) \\
 0 & 0 & U_C(\tau_u)
\end{matrix}
\right).
\end{equation}
Therefore, the above equation provides a way 
to compute the first- and second- order directional derivatives.

\subsection{Gradient-based Algorithm}
To maximize the fidelity and minimize the magnitudes of the directional derivatives, we define a   fitness function 
\begin{equation}
	\Phi(u) = \left| \operatorname{Tr}\left( U_C(\tau_u) R^\dag_0(\theta)\right) \right|^2 - \sum_{m=1}^{m_{\max}} \sum_{i_1,...,i_m =1}^2 \mu^{(m)}_{i_1,...,i_m} \left\| \mathcal{D}_{U_C}^{(m)}(V_{i_1},...V_{i_m}) \right\|^2,
\end{equation}
in which $m_{\max}$ is the maximum order of directional derivatives   involved, $\mu^{(m)}_{i_1,...,i_m} > 0$ are a set of weights specifying the importance of the associated objective function and are often   adjusted during optimization.  

The Gradient Ascent Pulse Engineering (GRAPE) algorithm
proposed in Ref. \cite{Khaneja05} is a standard gradient-based numerical method for solving  quantum optimal control problems. GRAPE works as follows
\begin{enumerate}[itemsep=0pt]
	\item[(1)] start from an initial pulse guess $u[l]$, $l=1,\ldots,L$;
	\item[(2)] compute the gradient of the fitness function with respect to the pulse parameters $\partial \Phi / \partial u[l]$;
	\item[(3)] determine a gradient-related search direction $p[l]$, $l=1,\ldots,L$;
	\item[(4)] determine an appropriate step length $\alpha$ along the search direction $p$;
	\item[(5)] update the pulse parameters: $u[l] \leftarrow u[l] + \alpha \times p[l]$;
	\item[(6)] if $\Phi$ is sufficiently high, terminate; else, go to step 2.
\end{enumerate}
 
\begin{figure*} 
	\includegraphics[width=0.9\linewidth]{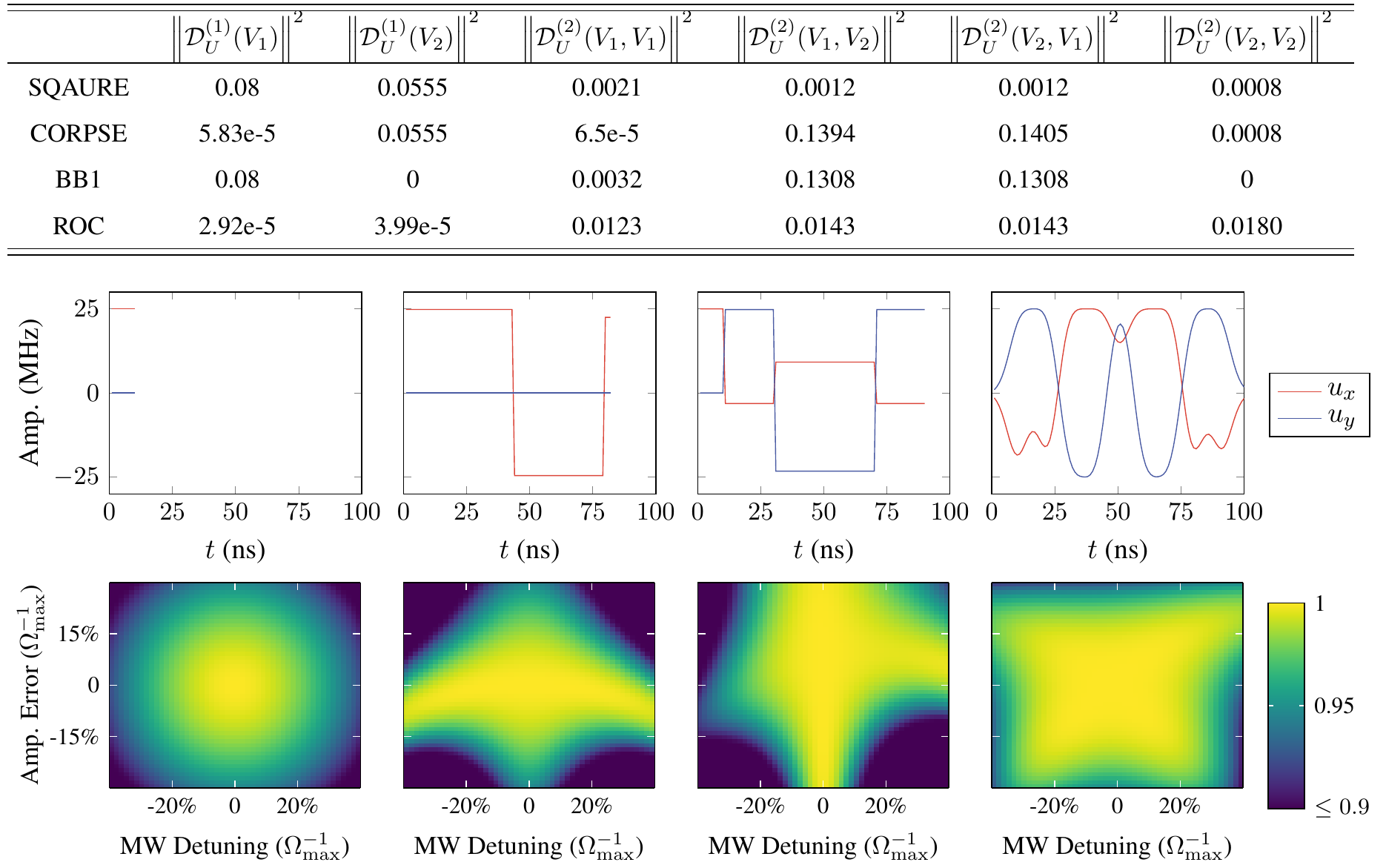}
	\caption{Results for $R_0(\pi/2)$. Comparison of: magnitudes of directional derivatives  up to second order, pulse shapes, and robustness profiles. From left to right are SQUARE, CORPSE, BB1, and ROC, respectively.}
		\label{pulse_90}
	\bigskip
	\includegraphics[width=0.9\linewidth]{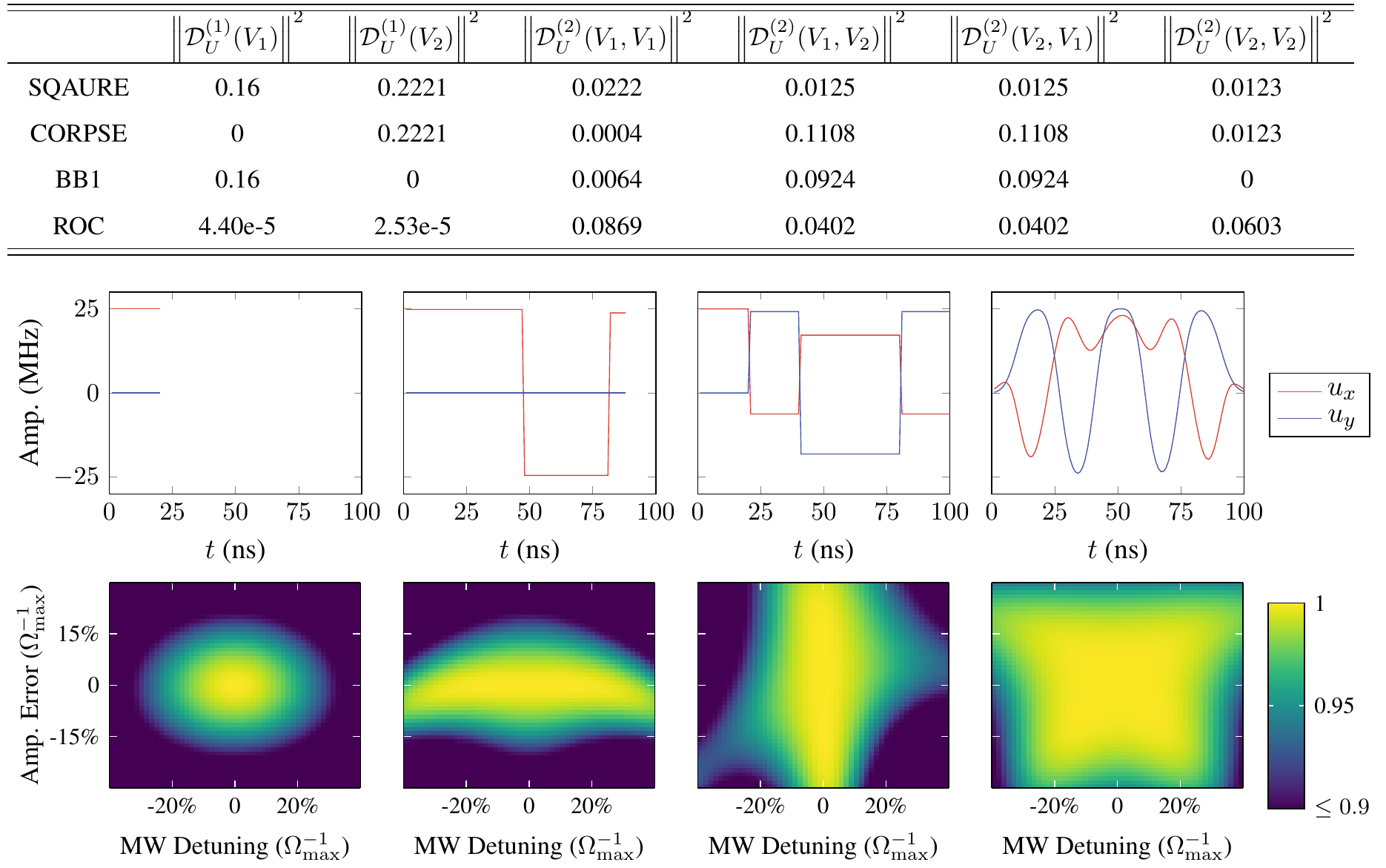}
	\caption{Results for $R_0(\pi)$. Comparison of: magnitudes of directional derivatives  up to second order, pulse shapes, and robustness profiles. From left to right are SQUARE, CORPSE, BB1, and ROC, respectively.}
		\label{pulse_180}
\end{figure*}

\subsection{Differential Evolution Algorithm}

As a powerful global optimization algorithm, differential evolution (DE) \cite{das2011differential} has found many successful applications in quantum control problems \cite{zfk2001,zahedinejad2015high,ZG16,YL19}. Here, we suggest DE as a promising candidate for seeking robust optimal pulses. The parameterized piecewise constant shaped pulse is usually expressed as a vector $u$, called individual in the DE context. The goal is then to maximize the above control fitness function $\Phi(u)$ over $u$. DE functions by performing mutation, crossover and selection operations in the population space which consists of a set of individuals. The algorithm procedure can be described as follows:

{Step 1}: Set the algorithm constants: scaling factor $R$, crossover rate $C_r$, chromosome length (the dimension of each individual) $p$ and population size $P_s$. Generate an initial population $P= \{  u_1,..., u_{P_s}\}$ randomly with $ u_i = (u_{i1},...,u_{ip})$ being the $i$-th individual in current population. 

{Step 2}: From $i=1$ to $P_s$, do the following steps: \begin{enumerate}[itemsep=0pt]
	\item[(1)] Mutation.  Generate a donor vector $v_i=(v_{i1},...,v_{ip})$ through the mutation rule: 
$
{ v_i} = { u_{r_b^i}} + R({ u_{r_1^i}} - { u_{r_2^i}}+{u_{r_3^i}} - { u_{r_4^i}}),	
$
 where $r_1^i, r_2^i, r_3^i, r_4^i$ are randomly chosen mutually exclusive integers in the range $[0, P_s]$ and $r_b^i$ is the index of the best individual in the current population.
 \item[(2)] Crossover. Generate a trial vector $\mu_i=(\mu_{i1},...,\mu_{ip})$ by a binomial crossover strategy: if $\text{rand}_{i,j}[0,1] \le C_r$ or $j=j_{\text{rand}}$, let $\mu_{ij}=v_{ij}$, where $j_{\text{rand}} \in \{1,2,...,p\}$ is a randomly chosen index. Otherwise let $\mu_{ij}=u_{ij}$.
 \item[(3)] Selection. Evaluate the former individual $u_i$ and the trail vector $\mu_i$, if $\Phi(\mu_i)\ge \Phi( u_i)$, let $ u_i=\mu_i$, otherwise keep $u_i$ unchanged.
\end{enumerate}

{Step 3}: Check the stopping criterion, if not satisfied, continue with {Step 2}.

\subsection{Results: Comparison with Composite Pulses}
Here, we compare the robustness to detuning errors and control amplitude errors between SQUARE, CORPSE, BB1, and our ROC pulses. CORPSE and BB1 are two commonly used composite pulses defined by:
\begin{align}
	\text{CORPSE: } & R_0(\theta) = R_0(\theta/2- \theta')R_\pi(2\pi-2\theta')R_0(2\pi + \theta/2 - \theta') \nonumber \\
	\text{BB1: } & R_0(\theta) = R_{\phi'}(\pi) R_{3\phi'}(2\pi) R_{\phi'}(\pi) R_{0}(\theta) \nonumber
\end{align}
where $\theta' = \arcsin [\sin(\theta/2)/2]$, $\phi' = \arccos(-\theta/4\pi)$
We focus on two rotational operations $R_0(\pi/2)$  and $R_0(\pi)$.  The results are presented in Fig. \ref{pulse_90} and in Fig. \ref{pulse_180}.

\section{Robust DD: Application Examples}

\subsection{Quantum Identity} 
The target is the identity operation. We compare between  a number of robust DD sequences that employ different basic pulses and different phase modulation strategies. Two   phase modulation strategies are considered here:
\begin{enumerate}[itemsep=0pt]
	\item The XY8 sequence XYXYYXYX.
	\item The universally robust (UR) DD Sequence proposed in Ref. \cite{GGT17}. UR DD sequence consists a sequel of  $n$ equally separated phased pulses. According to \cite{GGT17},   the phases of a UR sequence of $n$ pulses is given by
$$ \phi_k^{(n)} = \frac{(k-1)(k-2)}{2} \Phi^{(n)} + (k-1) \phi_2, $$
where
$ \Phi^{(4m)} = \pm  \pi/m$, and $\Phi^{(4m+2)} = \pm (2m\pi)/(2m+1)$. 
\end{enumerate}
The results are as follows:
\begin{center}
\includegraphics[width=0.9\linewidth]{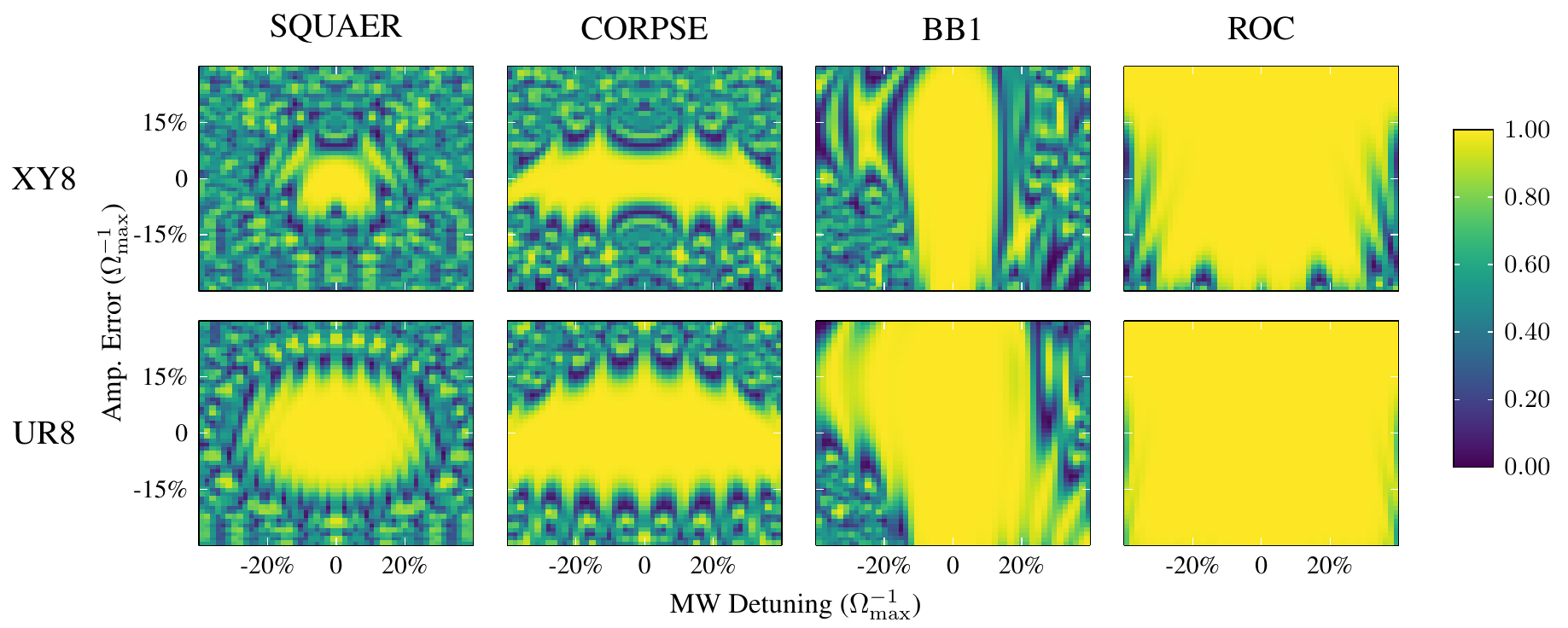}	
\end{center}	
Here, the control errors are measured in terms of the   maximum Rabi frequency $\Omega_{\max}$. All DD sequences contain $N= 100$ cycles and   have equal total time length of 3.2 ms. From the figure, we clearly see that
\begin{enumerate}
	\item[(1)] for the same basic pulse, UR8 outperforms XY8;
	\item[(2)] for the same phase modulation strategy, our ROC pulse performs best.
\end{enumerate}

\subsection{Detection of Nuclear Spins}

\subsubsection{Basic principles and representative detection  sequences}
In NV center system, through applying a subtle multipulse DD sequence on the electron spin, the weak signal of a specific nuclear spin can be amplified, meanwhile all the other nuclear spins can be  dynamically decoupled from the central electron spin. As such, we can resolve and coherently control the nuclear spins. Specifically, the electron spin is firstly prepared at $\rho=|+\rangle \langle+|$ with $|+\rangle=(|0\rangle+|1\rangle)/\sqrt{2}$ and a DD sequence then follows. Finally, the resultant electron spin state $\rho_t$ is attached a $R_{\pi/2}(\pi/2)$ rotation and measured in the basis $|0\rangle$, thus we can obtain the transition probability $P=\text{Tr}(\rho_f|0\rangle\langle0|)$. Explicitly, this procedure can be described as: 
\begin{equation}
|0\rangle\langle0|   \xRightarrow{R_{\pi/2}(\pi/2)} |+\rangle\langle+| \xRightarrow{\text{DD}} \rho_t  \xRightarrow{R_{\pi/2}(\pi/2)} \rho_f  ~{\Longrightarrow}~  \text{Tr}(\rho_f|0\rangle\langle0|). 
\end{equation}
Consider the basic 8-pulse DD unit in ideal case:
\begin{equation}
	\frac{\tau}{2} - R_{\phi_1}(\pi)- \tau - R_{\phi_2}(\pi)- \tau - R_{\phi_3}(\pi)- \tau - R_{\phi_4}(\pi)- \tau - R_{\phi_5}(\pi)- \tau - R_{\phi_6}(\pi)- \tau - R_{\phi_7}(\pi)- \tau - R_{\phi_8}(\pi)-\frac{\tau}{2},
\end{equation}
where $R_{\phi_k}(\pi)$ represents a $\pi$ rotation around the axis $S_{\phi_k}=\cos(\phi_k) S_x+ \sin(\phi_k) S_y$.
When the interpulse delay $\tau=k\pi/\omega_I$ ($k$ is odd number), the corresponding nuclear spin with effective Larmor frequency $\omega_I$ can induce a sharp resonant dip in the observed signal (time domain) \cite{TTW12}. This basic unit is usually cycled $L$ times to enhance the signal. The position and amplitude of the dip are then used to infer the hyperfine couplings between the electron spin and the sensed nuclear spin. Equivalently, we can transform this signal to frequency domain, i.e., a sharp resonant peak with detection frequency $\omega=\omega_I/k$, for clearly acquiring the spectrum information.

In realistic case, many common pulse imperfections can severely hinder the detection process. Finite width of the $\pi$ pulses, marked as $t_\pi$, will shift the resonant positions to $\tau^\prime=k\pi/\omega_I-t_\pi/2$. Detuning and pulse amplitude error can cause spurious peaks and distortion of the baseline. Thus, robust DD sequences that can resist these pulse imperfections are highly demand. In this work, we consider the following representative DD sequences ($\Omega_{\max}=25~$MHz). 

(1) XY8 sequence. The phases of the basic pulses are $\phi^{\text{XY8}}_k=\{ 0,\pi/2,0,\pi/2, \pi/2,0,\pi/2,0 \}$. The resonant positions are $\tau^\prime=k\pi/\omega_I-t_\pi/2$ with $t_\pi=20~$ns.

(2) UR8 sequence \cite{GGT17}. The phases of the basic pulses are $\phi_k^{\text{UR8}}=\{ 0,\pi/2,3\pi/2,\pi, \pi,3\pi/2, \pi/2,0 \}$. The resonant positions are $\tau^\prime=k\pi/\omega_I-t_\pi/2$ with $t_\pi=20~$ns.

(3) BB1-XY8 sequence \cite{BKR04}. The phases of the basic pulses are $\phi^{\text{BB1-XY8}}_k=\{ 0,\pi/2,0,\pi/2, \pi/2,0,\pi/2,0 \}$. Each $R_{\phi_k}(\pi)$ is realized by the BB1 composite pulses, namely $R_{\phi_k}(\pi)=\Pi_{i=1}^4 R_{\phi_i}(\theta_i)$ with $\theta_1=\theta_3=\pi,\theta_2=2\pi,\theta_4=\pi;\phi_1=\phi_3=\phi_k+\text{accos}(-1/4),\phi_2=3\phi_1-2\phi_k,\phi_4=\phi_k$. The resonant positions then become $\tau^\prime=k\pi/\omega_I-t_\text{BB1}/2$ with $t_\text{BB1}=5t_\pi=100~$ns.

(4) RP-XY8 sequence \cite{WZL19}. The phases of the basic pulses are $\phi_k^{\text{RP-XY8}}=\phi^{\text{XY8}}_k+\phi_n, \phi_n \in \text{rand}[0,2\pi], n=1,2,...,N$. That is, the 8 basic pulses in each unit are attached a random global phase. The resonant positions are $\tau^\prime=k\pi/\omega_I-t_\pi/2$ with $t_\pi=20~$ns.

(5) GAXY8 sequence \cite{HJW18}. The phases of the basic pulses are $\phi^{\text{GAXY8}}_k=\{ 0,\pi/2,0,\pi/2, \pi/2,0,\pi/2,0 \}$. Each basic $\pi$ rotation is then realized by the 5-pulse composite AXY8 sequence \cite{CJW15}. Further, each AXY8 sequence employs a different amplitude but keeps the same in a DD unit. In this way, the soft modulation can be resembled as the Gaussian shape $\lambda(t)=\lambda_{0} \exp \left[-t^{2} /\left(2 \sigma^{2}\right)\right]$ by the sequence of hard $\pi$ pulses, we set $\sigma=T/(4\sqrt{2})$ with $\lambda_0=0.271,T=4L\tau$. The resonant positions are $\tau^\prime=k\pi/\omega_I-t_\text{AXY8}$ with $t_\text{AXY8}=5t_\pi=100~$ns.

\subsubsection{Searching robust optimal pulses for improving DD robustness}
As mentioned above, a multipulse DD sequence can select the signal of a specific nuclear spin, meanwhile dynamically decouple the other nuclear spins. Thus, a system consists of the central electron spin and one target nuclear spin is sufficient to characterize the issue in multi-spin case. For designing robust optimal pulses, we consider the commonly encountered pulse imperfections, including detuning, amplitude error, and finite width of $\pi$ and $\pi/2$ rotations. These factors are treated as perturbations, resulting in the perturbed Hamiltonian 
  \begin{equation}
	H(t) =\Omega_I I_z+A_{zz}S_z I_z+A_{zx}S_x I_z+(1+\epsilon_\Omega)\Omega(t)[S_x \cos(\varphi(t))+S_y \sin(\varphi(t))]+\epsilon_z \Delta_{\max} S_z. 
\end{equation}
Compared to its general form $H(t)=H_{S E}+H_{C}(u(t))+\sum_{i} \epsilon_{i} V_{i}(t)$, we have $H_{SE}=\Omega_I I_z+A_{zz}S_z I_z+A_{zx}S_x I_z, H_{C}(u(t)) =\Omega(t)[S_x \cos(\varphi(t))+S_y \sin(\varphi(t))], V_1= \Delta_{\max}S_z, V_2(t)=\Omega(t)[S_x \cos(\varphi(t))+S_y \sin(\varphi(t))],\varepsilon_1=\epsilon_z,\varepsilon_2=\epsilon_\Omega$. 
In our simulations, we concern the case with relatively large pulse imperfections, thus set $\varepsilon_z=\varepsilon_\Omega=0.08,\Delta_{\max}=\Omega_{\max}=25~$MHz. Besides, though we do not know the hyperfine couplings before detection, we can bound them by certain upper values. As such, we set $A_{zz}=A_{zx}\leq 0.1\Omega_I$ with $\Omega_I=\gamma_I B, \gamma_I=2\pi \times 1.071~$kHz/G, $B=400~$G. 

\begin{figure}
\centering
\includegraphics[width=0.8\linewidth]{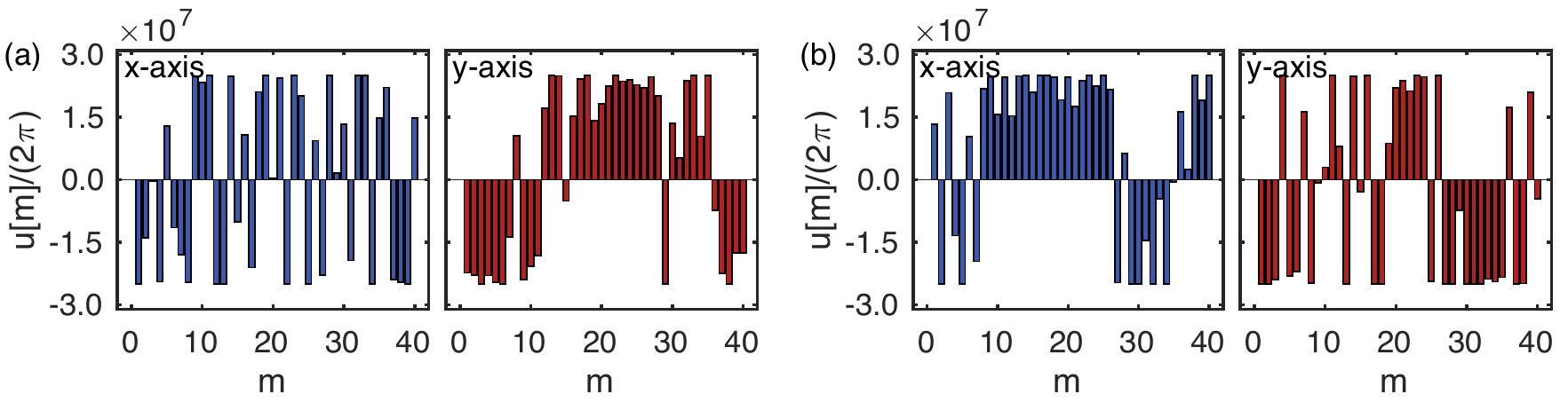}	
\caption{The robust optimal pulses searched by differential evolution algorithm. The pulses are divided into 40 piecewise-constant slices along $x$ and $y$ axes with a total pulse length being $80~$ns. (a) The pulses for $R_{\pi/2}(\pi/2)$ rotation. (b) The pulses for $R_{0}(\pi)$ rotation.}
\label{rop}
\end{figure}

In convenience, the time-varied controls $u(t)=(u_x(t),u_y(t))=(\Omega(t)\cos(\varphi(t)),\Omega(t)\sin(\varphi(t))):t\in[0,\tau_u]$ are divided into $M$ piecewise constant slices with equal duration time $\Delta t= \tau_u/M$. This transforms the controls into the form $\bm u=(u[m])=(u_x[m],u_y[m]):m\in[1,M]$. For a $\theta$ rotation with a phase shift $\phi$ in DD sequence, i.e., $R_\phi(\theta)=e^{-i \theta [\cos(\phi) S_x+ \sin(\phi) S_y]}$, the actual evolution operator can be calculated by 
\begin{equation}
	U_{\theta,\phi}=\Pi_{m=1}^M e^{-i \Delta t \{\Omega_I I_z+A_{zz}S_z I_z+A_{zx}S_x I_z+(1+\epsilon_\Omega)[u_x[m]S_x +u_y[m]S_y ]+\epsilon_z \Delta_{\max} S_z \} }.
\end{equation}
By denoting the main physical goal $f_0=\text{Tr}(U_{\theta,\phi} R^\dag_\phi(\theta))$ and the norm of the corresponding directional derivatives $f_1=||\mathcal{D}^{(1)}_{U_{\theta,\phi}}(V_1)||^2, f_2=||\mathcal{D}^{(1)}_{U_{\theta,\phi}}(V_2)||^2,f_3=||\mathcal{D}^{(1)}_{U_{\theta,\phi}}(H_{SE})||^2,f_4=||\mathcal{D}^{(2)}_{U_{\theta,\phi}}(V_1)||^2, f_5=||\mathcal{D}^{(2)}_{U_{\theta,\phi}}(V_2)||^2,f_6=||\mathcal{D}^{(2)}_{U_{\theta,\phi}}(H_{SE})||^2$, the multi-objective control fitness function can be formalized as $\Psi=p_0f_0+\sum_{i=1}^6 p_i(1-\alpha_i f_i)$. The parameters are carefully tuned and chosen as $p_0=1/2,p_i=1/12,\alpha_1=10^2,\alpha_2=\alpha_3=10^3,\alpha_4=10^4,\alpha_5=\alpha_6=10^5$. According to the ROC procedure using DE algorithm (the algorithm parameters are set as $R=0.6,C_r=0.95,p=80,P_s=60$, the optimization stops when the control performance function reaches 0.999 within 1000 iterations), the robust optimal pulses can be found, as shown in Fig. \ref{rop}. Incorporating these robust pulses into the investigated DD sequences can substantially enhance their robustness, as analyzed in the main text.

\subsubsection{Simulation details}
We supplement some details and explanations of Fig. 2 in the main text. 

\emph{Single spin.--}The hyperfine couplings and the effective Larmor frequency for the investigated single spin (Fig. 2(a) in the main text) are $A_{zz}^1=2\pi \times 27~$kHz,
$A_{zx}^1=2\pi \times 17~$kHz, $\omega_I=2\pi \times 428.41~$kHz. As analyzed, the $k$th order resonant peak appears at location $\omega_I/k$ with $k$ being odd. For brevity, we only show the resonant peaks up to third order.

\emph{Six spins.--}The detailed coupling constants between the NV electron spin and the surrounding six nuclear spins (Fig. 2(b) in the main text) are listed in Table. \ref{couplings}. For brevity, we look over the spectra range that contains all the eighth order resonant peaks of the six spins. It worth noting that when the cycle number $L$ is much larger, the spectrum using the XY8 sequence will be more miscellaneous, while combing our robust optimal pulses can still resist the considered pulse imperfections well.  

\begin{table}[h]
{\renewcommand{\arraystretch}{1.3}\setlength{\tabcolsep}{8pt}
\begin{tabular}{|c|ccc|c|ccc|}
  \hline
  Spin &~~$A_{zz}^n$ (kHz)~~ & ~~$A_{zx}^n$ (kHz)~~ & ~~$\omega^n_I$ (kHz)~~ & Spin &~~$A_{zz}^n$ (kHz)~~ & ~~$A_{zx}^n$ (kHz)~~ & ~~$\omega^n_I$ (kHz)~~\\
  \hline
  ~~1~~ & $2\pi \times 78.234$ & $2\pi \times 30.031$ & $2\pi \times 468.59$ & ~~4~~ & -$2\pi \times 12.958$ & $2\pi \times 13.896$ & $2\pi \times 422.99$\\
  ~~2~~ & $2\pi \times 40.703$ & $2\pi \times 23.500$ & $2\pi \times 449.82$ & ~~5~~ & -$2\pi \times 22.081$ & $2\pi \times 24.525$ & $2\pi \times 418.43$\\
  ~~3~~ & $2\pi \times 32.328$ & $2\pi \times 44.496$ & $2\pi \times 445.64$ & ~~6~~ & $2\pi \times 15.796$ & $2\pi \times 19.506$ & $2\pi \times 437.37$\\   
  \hline
\end{tabular}
\caption{The hyperfine couplings and the effective Larmor frequency for the six nuclear spins derived from Ref. \cite{TTW12}. The Larmor frequency for each nuclear spin is $\Omega_I=\gamma_I B$ with $\gamma_I=2\pi \times 1.071~$kHz/G and $B=401~$G. } \label{couplings}
}
\end{table}

\emph{Two close spins.--}In Fig. 2(c) in the main text, we use two spectrally close spins with couplings
$A_{zz}^1=2\pi \times 27.02~$kHz, $A_{zx}^1=2\pi \times 16.91~$kHz, $\omega_I^1=2\pi \times 441.91~$kHz
$A_{zz}^2=2\pi \times 18.28~$kHz, $A_{zx}^2=2\pi \times 54.26~$kHz, and $\omega_I^2=2\pi \times 437.53~$kHz. 
As described above, we use the optimal pulses that can minimize the norm of the directional derivatives with respect to the considered imperfections up to the second order. When applying higher order minimizations, we anticipate that the visible gap between our results and the ideal spectrum can be gradually bridged.

\end{document}